\documentclass[journal=jacsat,layout=twocolumn]{achemso}

\usepackage{graphicx}
\usepackage{amsmath}
\usepackage{amssymb}
\usepackage{xcolor}
\usepackage{enumitem}

\newcommand{\hw}{D$_2$O}

\newcommand{\nabph}{NaBPh$_4$}

\newcommand{\bphm}{BPh$_4^-$}
\newcommand{\mepy}{3-MP}
\newcommand{\NA}{Na$^+$}

\title{Monolayer Structure of Supramolecular Antagonistic Salt Aggregates}
\author{David Jung}
\affiliation{Helmholtz Institute Erlangen-N\"{u}rnberg for Renewable Energy, Forschungszentrum J\"{u}lich, F\"{u}rther Stra{\ss}e 248, 90429 N\"{u}rnberg, Germany.}
\author{Jens Harting}
\affiliation{Helmholtz Institute Erlangen-N\"{u}rnberg for Renewable Energy, Forschungszentrum J\"{u}lich, F\"{u}rther Stra{\ss}e 248, 90429 N\"{u}rnberg, Germany.}
\alsoaffiliation{Department of Chemical and Biological Engineering and Department of Physics, Friedrich-Alexander-Universit\"at Erlangen-N\"urnberg, F\"{u}rther Stra{\ss}e 248, 90429 N\"{u}rnberg, Germany.}
\email{j.harting@fz-juelich.de}
\author{Marcello Sega}
\affiliation{Helmholtz Institute Erlangen-N\"{u}rnberg for Renewable Energy, Forschungszentrum J\"{u}lich, F\"{u}rther Stra{\ss}e 248, 90429 N\"{u}rnberg, Germany.}
\email{m.sega@fz-juelich.de}
\phone{+49 911 32169-102}

\begin{document}
\begin{abstract}
	The speculated presence of monomolecular lamellae of antagonistic salts in oil-water mixtures has left several open questions besides their hypothetical existence, including their microscopic structure and stabilization mechanism. Here, we simulate the spontaneous formation of supramolecular aggregates of the antagonistic salt sodium tetraphenylborate (\nabph{}) in water and 3-methylpyridine (\mepy{}) at the atomistic level. We show that, indeed, the lamellae are formed by a monomolecular layer of the anion, enveloped by \mepy{} and hydrated sodium counterions. To understand which thermodynamic forces drive the aggregation, we compare the full-atomistic model with a simplified one for the salt and show that the strong hydrophobic effect granted by the large excluded volume of the anion, together with electrostatic repulsion, suffice to explain the stability of the monomolecular lamellae.
\end{abstract}
\maketitle

\section{Introduction}
\begin{figure}
  \centering
  \includegraphics[width=0.8\columnwidth]{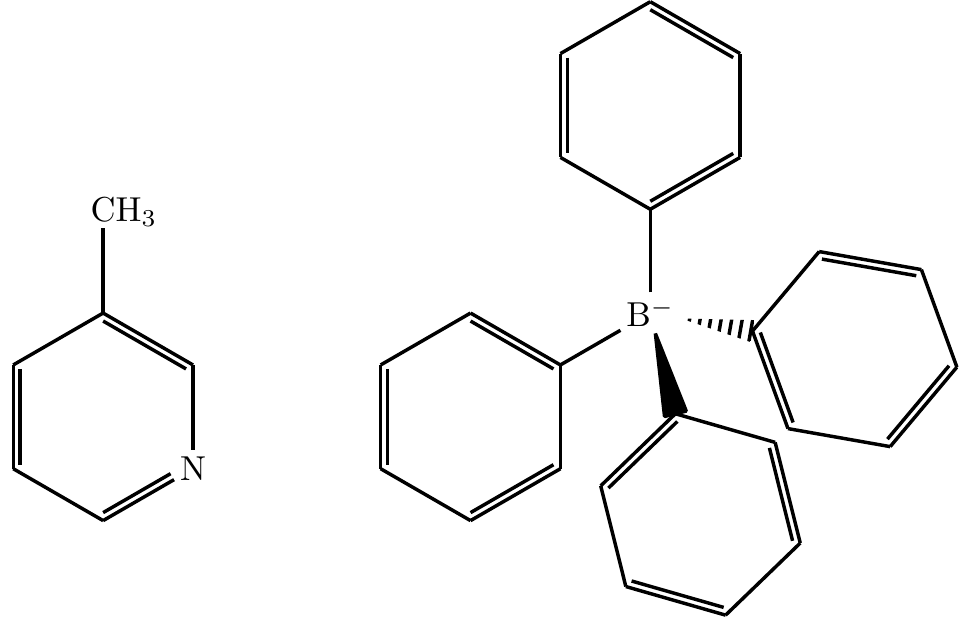}
  \caption{Structure formulae of \mepy{} (left) and \bphm{} (right).}
  \label{fig:struct_form}
\end{figure}

Antagonistic salts are considered a new class of surface-active solutes, with distinct features from the other two known classes, salts and surfactants\cite{Bonn2014}. Antagonistic salts consist of a relatively large organic ion and a small inorganic counterion. Unlike typical ionic surfactants, that dissociate into an amphiphilic ion and a small inorganic counterion, the organic ion of an antagonistic salt has a marked preference for organic solvents. In an oil/water mixture, the organic ions are solvated by the oil phase and the inorganic counterions by water, giving rise to the eponymous antagonistic behavior. A typical example of an antagonistic salt and oil mixture is that of \nabph{} and \mepy{}, as shown in Fig.~\ref{fig:struct_form}. Thanks to the electrostatic attraction between the organic anion and its counterions, an electric double layer forms at the water/oil interface, providing the surfactant property that manifests itself also by the appearance of typical mesostructures like lamellae\cite{sadakane_long-range_2006}. Oppositely to (homoselective) salts, antagonistic salts also widen the intermiscibility range of their solvent mixtures\cite{sadakane_long-range_2006,glende_vanishing_2020}. Interestingly, theoretical models by Onuki and Kitamura\cite{onuki_solvation_2004} predicted the appearance of antagonistic salt mesostructures before their experimental observation. Onuki and coworkers' theoretical approach is based on a Ginzburg-Landau mean-field model, which predicts the characteristic wave-like structures as a result of the coupling between solvent and charge concentration\cite{onuki_solvation_2004,onuki_phase_2011, onuki_structure_2016}. However, a later analysis showed that the theoretical predictions would fit small-angle neutron scattering (SANS) spectra only using an unrealistically large Debye length. A better fit to the experimental data could be obtained by using general scattering functions for lamellar structures\cite{sadakane_mesoscopic_2007}.

In successive experiments, Sadakane and coworkers studied mixtures of \nabph{} in \hw{}/\mepy{} and showed that at salt concentrations larger than 50~mM the two-phase region of \hw{} and \mepy{} shrinks up to a point where no macroscopic phase separation can be detected at all.\cite{sadakane_periodic_2007} Remarkably, the structures formed by \nabph{} reached characteristic lengths close to the wavelength of visible light when the temperature was within a few degrees of the critical point, leading to the observation of mixture coloration that changes sensitively with temperature\cite{sadakane_periodic_2007}.
SANS and neutron spin echo measurements confirmed the presence of a disordered phase and, in addition, of an ordered multilamellar vesicle phase  at \mepy{} volume fractions less than $ 11\%$. The thickness of the fluctuating lamellae turned out, surprisingly, to be comparable to the size of the composing \bphm{} anions\cite{sadakane_membrane_2013}.
The formation of periodic mesostructures by antagonistic salts has also been confirmed in slightly different solvent mixtures, such as \hw{} and dimethylpyridine,\cite{sadakane_lamellar_disorder_2014} and using other salts with similar ionic size asymmetry, such as tetraphenylphosphonium chloride.\cite{sadakane_effect_2012}

Several mean-field level theoretical and numerical studies have so far delivered some insight into the phase diagram and structure size dependencies of the supramolecular aggregates\cite{araki_dynamics_2009,tasios_microphase_2017,jung_how_2019}. In these mean-field studies, the driving force segregating the antagonistic ions into opposite fluid phases is derived from a phenomenological solvation potential varying linearly with the fluid composition. However, little is known so far of the aggregate structure at the molecular scale and about the mechanisms driving mesostructure formation in these mixtures.
Motivated by the surprisingly thin, molecular-scale lamellae suggested by the experiments in the ordered phase, which current theoretical models cannot consistently explain, we set out to determine how these lamellar structures form and what stabilizes them by simulating both an atomistic model of \nabph{}/\mepy{}/water ternary mixtures, as well as a mixed atomistic/coarse-grained bead model where each \bphm{} is represented by a single charged interaction point interacting purely repulsively with all others.
\section{Methods}
Ionic solutions are notoriously difficult to simulate with empirical potentials, where the neglect of charge transfer and electronic polarizability often results in excessive ion pairing and too low solubilities\cite{smith2018recent}. Charge scaling has proven to be a simple but effective approach to improve the reliability of empirical potentials for ionic systems. Therefore, we modeled sodium ions using the Madrid model, which rescales the charges of anions and cations by a factor of 0.85\cite{benavides_potential_2017}, in combination with the TIP4P/2005 interaction potential for water\cite{abascal_general_2005}, which was used to parametrize the Madrid model. As for the empirical potentials for \bphm{},  we started from a template generated by ATB\cite{malde_automated_2011} using density functional calculations at the B3LYP/6-31G* level of the theory\cite{becke1993_3,lee1988development,perdew1992accurate} and the polarizable continuum implicit solvent model\cite{miertuvs1981electrostatic}. Partial charges were assigned using the Merz-Singh-Kollman scheme\cite{singh1984approach} and scaled by a factor of 0.85 to be consistent with the Madrid model and maintain charge neutrality. We added distance constraints to keep the structure of the molecule rigid, and used the all-atom optimized potential for liquid simulation (OPLS/AA) Lennard-Jones parameters\cite{jorgensen2005potential} as reported in the SI.

As a starting template for \mepy{}, we used the topology that Caleman and coworkers used in their benchmark of organic liquid force fields\cite{caleman_force_2012}, based on OPLS/AA. The thermodynamic properties of the water/\mepy{} mixture simulated using empirical potentials are known to depend on a delicate balance between the solvation of nitrogen and the dispersion interaction with other \mepy{} molecules\cite{almasy_qens_2002}. Experimentally, \mepy{} and water mix in every proportion\cite{andon_896._1952}, while heavy water and \mepy{} produce a closed-loop phase diagram\cite{cox_897._1952}. Our preliminary simulations of water/\mepy{} mixtures using OPLS/AA and TIP4P/2005 revealed an unphysical phase separation at 9\% \mepy{} volume fraction and all tested temperatures from 250 to 340~K. To improve the quality of the \mepy{} model we followed the force field calibration protocol of Salas and coworkers\cite{salas_systematic_2015} by changing the Lennard-Jones parameters and rescaling the original OPLS/AA partial charges of \mepy{} atoms until we matched within 7\% \mepy{} dielectric permittivity\cite{madelung_static_1991}, density\cite{haynes_crc_2016}, and liquid/vapor surface tension\cite{lechner_surface_2016} at 300~K. The new parameterization resulted in a single-phase mixture of water/\mepy{} in the whole tested range of \mepy{} concentrations and temperatures ranging from 10-50~vol\% and 275-350~K respectively, as expected for light water.
The GROMACS topology files, as well as equilibrated configurations, are available at the central institutional repository for research data of Forschungszentrum J\"ulich, DOI \texttt{10.26165/JUELICH-DATA/OWA443}\cite{JUELICH-DATA-OWA443}.

All our simulations were performed in the isobaric-isothermal ensemble using the Parrinello-Rahman pressure coupling\cite{parrinello_polymorphic_1981} and employing the Nos\'e-Hoover\cite{nose_molecular_1984,hoover_canonical_1985} thermostat with relaxation constants of 1.2 and 0.6~ps, respectively.
For the fully-atomistic simulations, we took into account the long-range electrostatics and dispersion corrections using the smooth variant of the Particle Mesh Ewald method\cite{essmann_smooth_1995} with a grid spacing of 0.12~nm, cubic interpolation scheme, and $10^{-5}$ and $10^{-4}$ as relative contribution of the direct part of the electrostatic and Lennard-Jones potential at the cutoff.  For the coarse-grained model of \bphm{}, we computed the Lennard-Jones interaction up to a cutoff distance of $2^{1/6}$~nm.

All simulations were initialized in a cubic simulation cell of either 10 or 24~nm initial side length by randomly placing (taking care of avoiding overlaps) the ions, \mepy{}, and eventually water molecules. After an initial energy minimization using the steepest descent algorithm, the systems underwent a preliminary relaxation of 50~ps at a reference pressure of 500~bar to speed up condensation. The systems were further relaxed at a pressure of 1~bar, where the simulation cell edges oscillated around approximately 8 and 20~nm respectively.
All simulations were performed using GROMACS 2019\cite{abraham_gromacs_2015} with an integration step of 2~fs, for a minimum of 50~ns, dumping configurations to disk every 500~ps for subsequent analysis.
\section{Results}
\subsection{Local structure of binary and ternary mixtures}
\begin{figure*}
  \centering
 \includegraphics[width=0.3\textwidth]{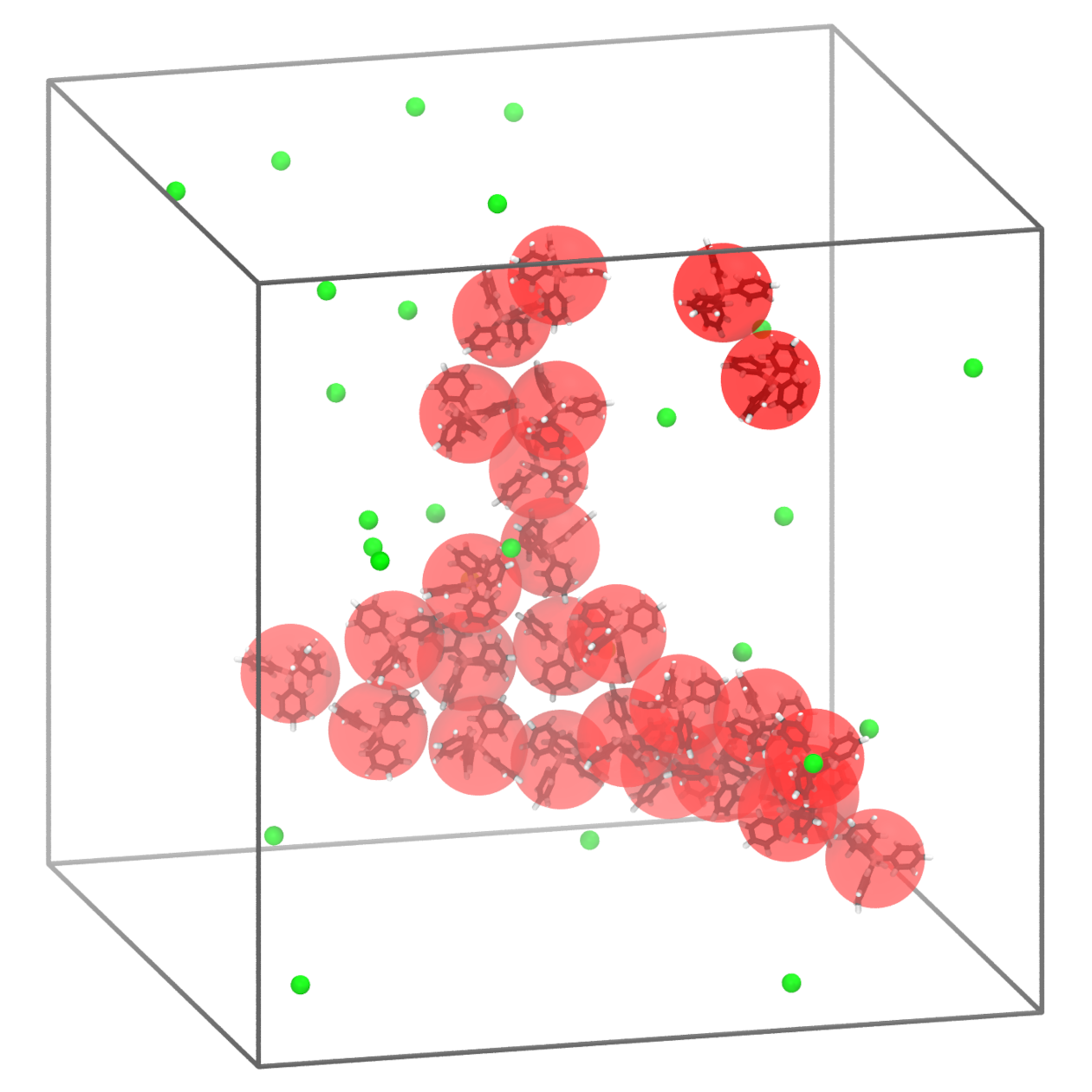}
 \includegraphics[width=0.3\textwidth]{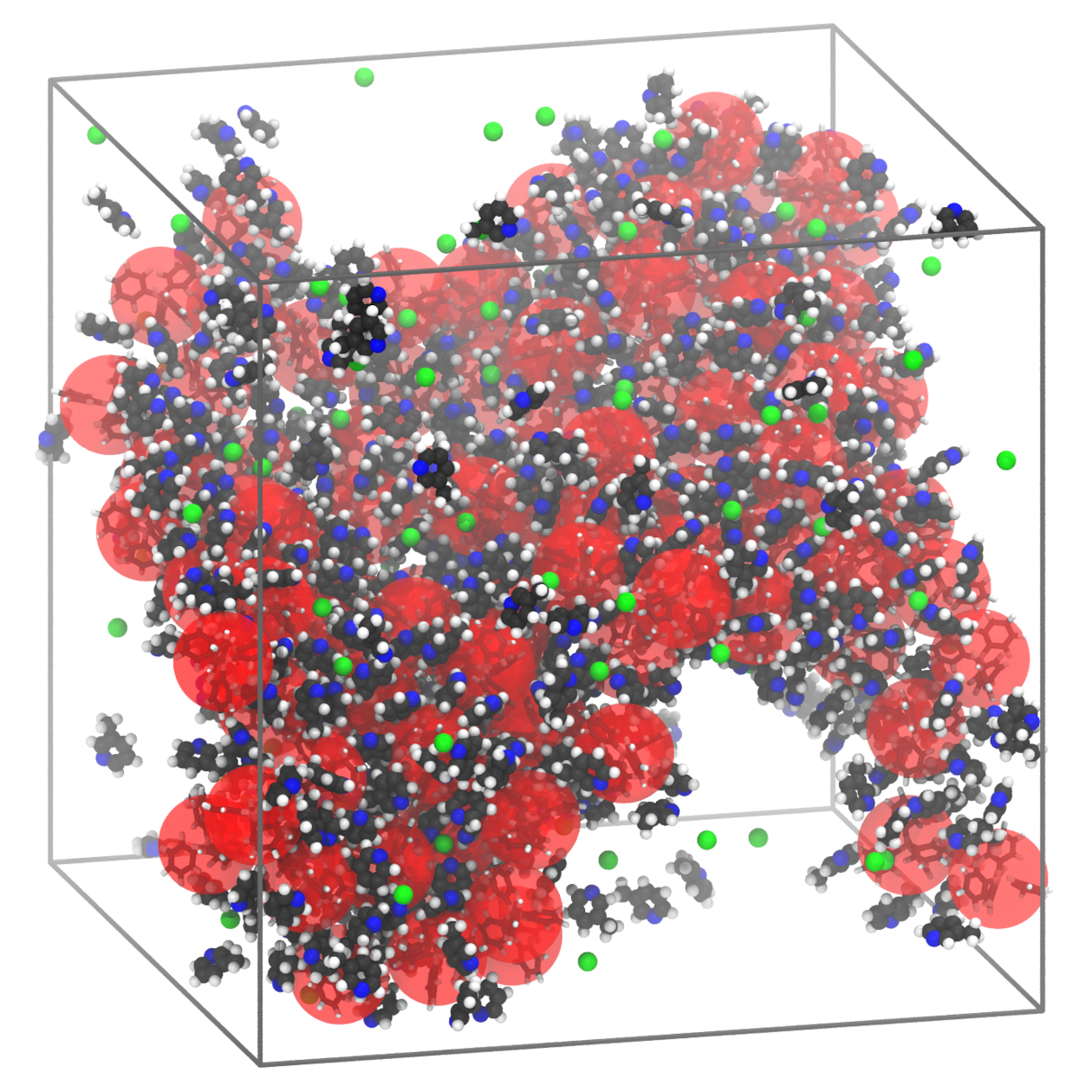}
 \includegraphics[width=0.3\textwidth]{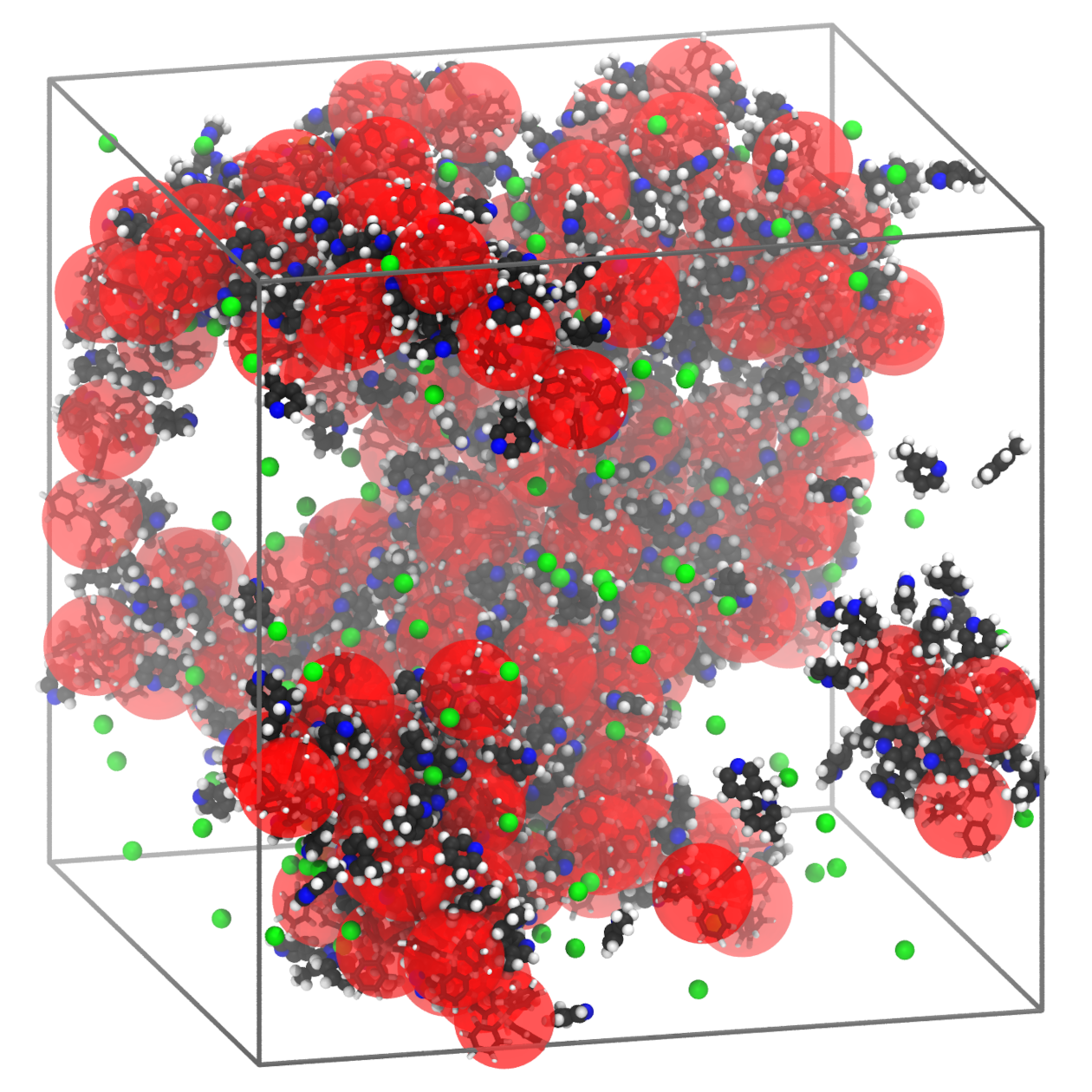}
 \includegraphics[width=0.3\textwidth]{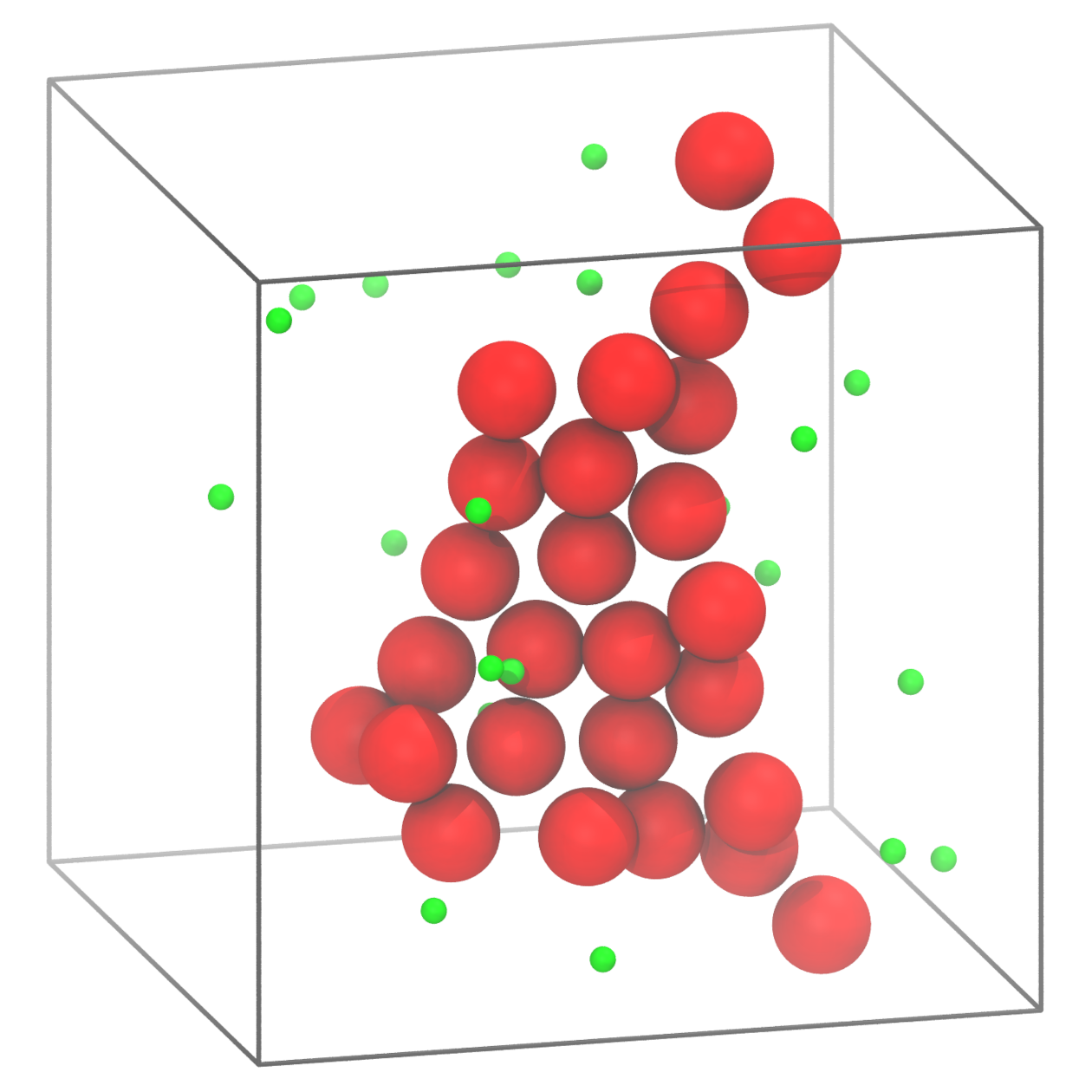}
 \includegraphics[width=0.3\textwidth]{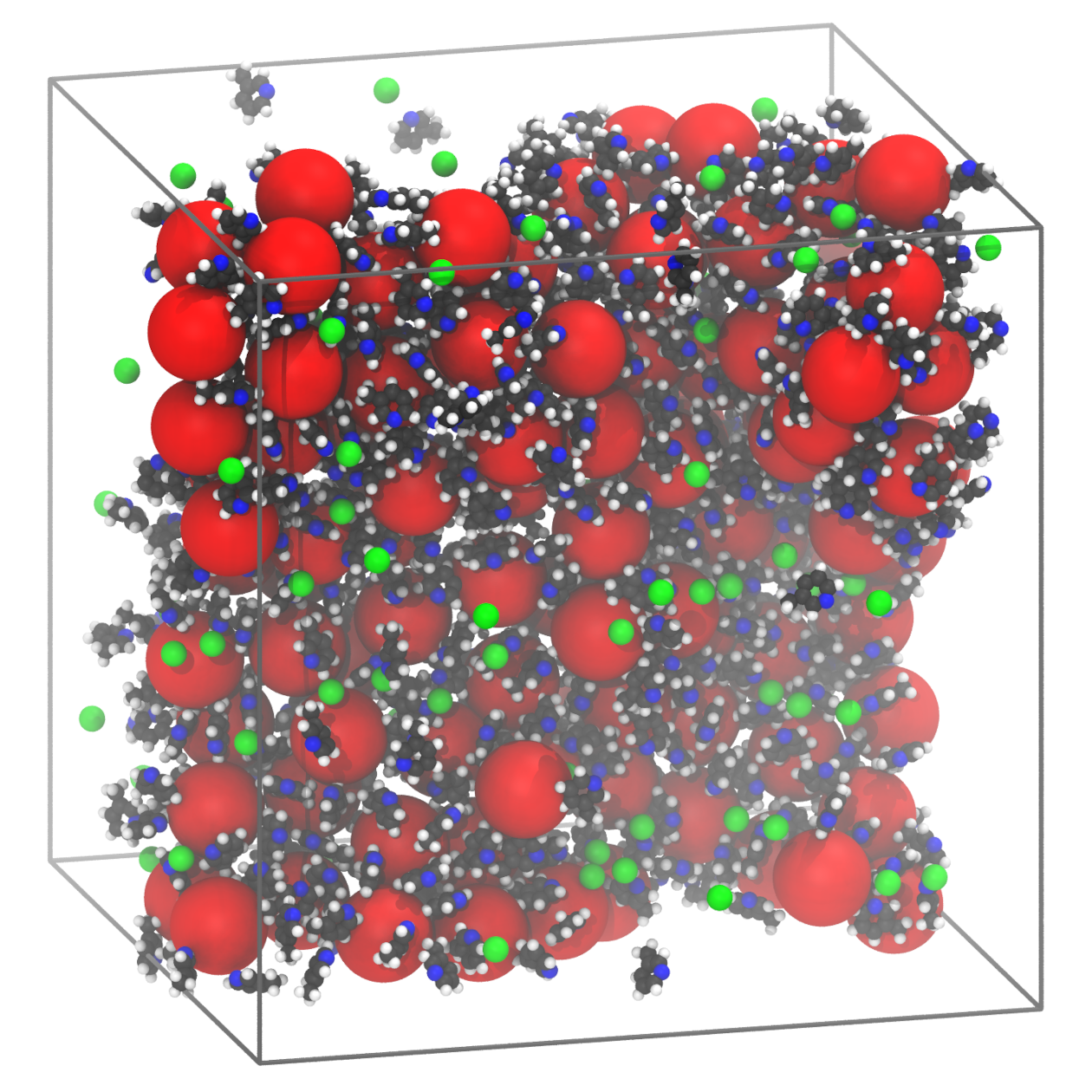}
 \includegraphics[width=0.3\textwidth]{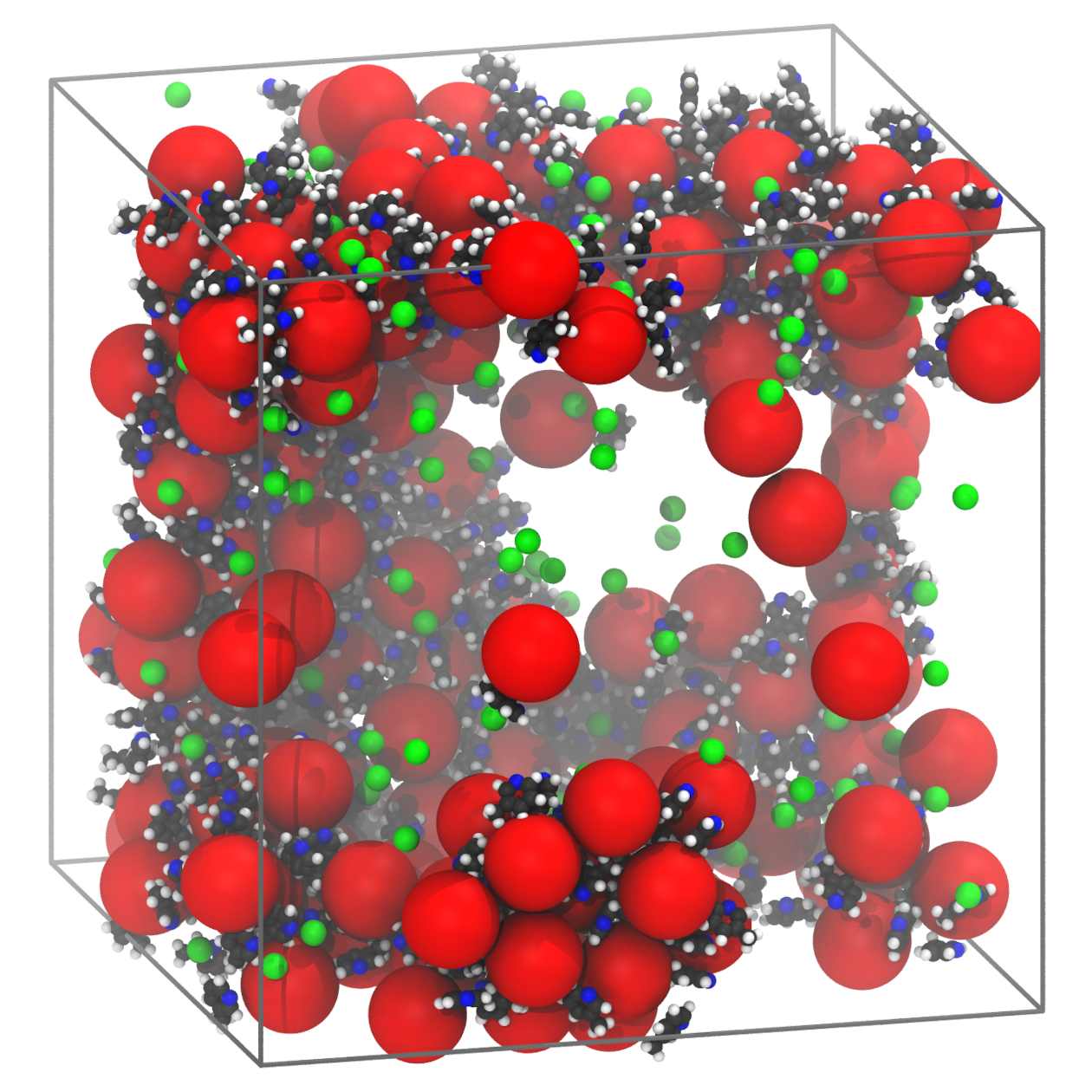}
 \caption{Simulation snapshots of the atomistic (top row) and bead model (bottom row) of \nabph{}/\mepy{}/water simulations at $T=300$K, 15~vol-\% \mepy{} and a \nabph{} concentration of 80 (left), 300 (center)  and 500~mM (right). The snapshots show \NA{} (green spheres), \bphm{} (sticks representation plus red halo in the atomistic case; red spheres in the bead model case)  and \mepy{} (balls-and-stick representation). For clarity, water molecules are not shown in any snapshot, and \mepy{} is not shown in the 80~mM case.}
  \label{fig:bead_sample}
\end{figure*}

In Fig.~\ref{fig:bead_sample}, we show some snapshots from equilibrated simulations of the ternary solution at different salt concentrations. All snapshots show the full simulation system about 50~ns after initialization from a fully randomly mixed state different for each simulation. The typical structures observed are planar ones in the form of leaflets at low salt concentrations, 2D-spanning membranes at medium salt concentrations, and 3D percolating, bicontinuous structures at high salt content.

Before detailing the composition and structure of these aggregates, it is helpful to understand the role of the different components in the simpler \bphm{}/\mepy{} and \bphm{}/water binary mixtures.
\begin{figure}
  \centering
  \includegraphics[width=\columnwidth]{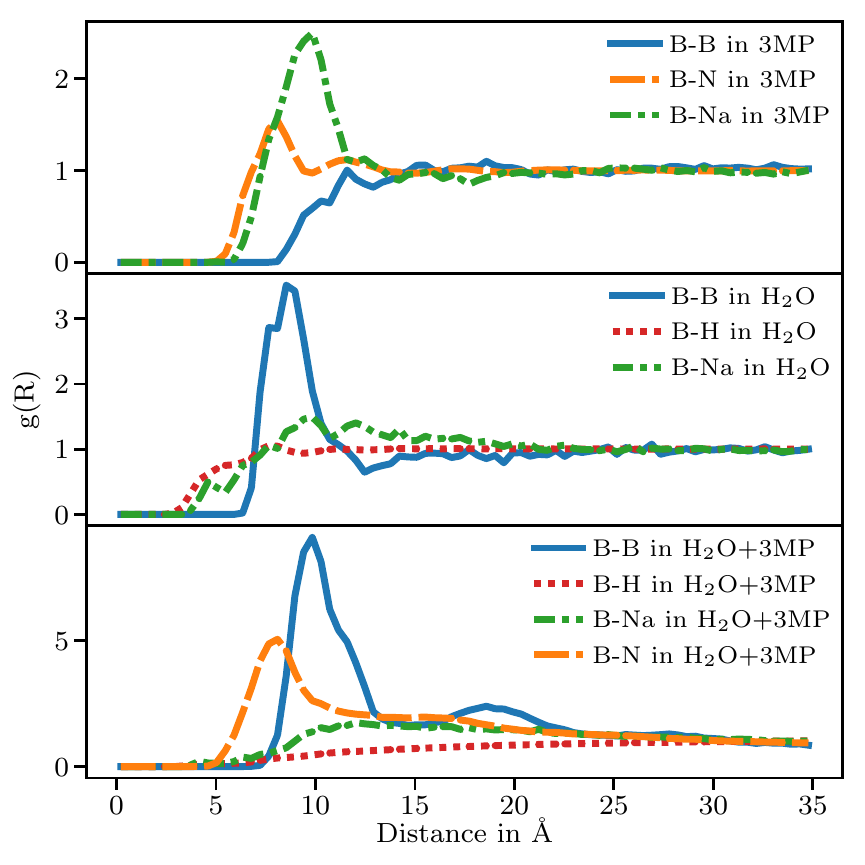}
  \caption{RDFs of 200~mM \nabph{} solutions in pure \mepy{} (top), pure water (center), and ternary mixture with 15~vol\% \mepy{} (bottom). N is the nitrogen in \mepy{} and H is the hydrogen in water.}
  \label{fig:rdf}
\end{figure}

In \nabph{}/\mepy{} mixtures the presence of \mepy{} is sufficient to fully solvate \nabph{}, as a quick look at the boron radial distribution function (RDF) in the top panel of Fig.~\ref{fig:rdf} confirms, although \bphm{} and \NA{} remain partially associated, with a coordination number of 1.65. \mepy{} clearly forms up to two solvation shells around \bphm{}, as indicated by the dashed curve in the upper panel of Fig.~\ref{fig:rdf}.

On the other hand, \nabph{}/water mixtures at low to moderate salt concentrations show the appearance of small, transient structures, mostly pairs or short chains of three \bphm{} ions. These clusters are responsible for the peak in the boron RDF shown in the central panel of Fig.~\ref{fig:rdf}.  In this sense, water appears to be a worse solvent for \nabph{} than \mepy{}. This is somewhat surprising because \mepy{} has a noticeably weaker dipole moment than water. However, one needs to consider the fact that water can fully dissociate the \nabph{} into solvated ion pairs.
In contrast, a significant number of ion pairs persist in \mepy{}, as shown by the coordination number. The formation of transient pairs and short chains of \bphm{} can be attributed to the hydrophobic character of the large \bphm{} ion, where the charge of boron is buried behind the four phenyl rings.

When the ternary mixture is considered, the boron RDF changes dramatically, showing a much more prominent primary peak and a sizeable secondary peak. Together with the slow decay of all correlation functions, these two peaks suggest the presence of extended clusters with a dimensionality lower than 3. From the detail of the cross-sectional structure shown in Fig.~\ref{fig:profile} as well as from the boron-nitrogen RDF in the bottom panel of Fig.~\ref{fig:rdf}, it is clear that \mepy{} envelops the \bphm{} aggregates with one molecular layer. Here \bphm{} displays hydrotropic behavior, since it aggregates much more than in either pure water or pure \mepy{}.

The primary peak of the boron RDF shifts by 0.1~nm to larger distances as compared to the case of water only. One could argue that due to the presence of surrounding \mepy{}, the charges of \bphm{} interact effectively through a medium with lower dielectric permittivity than water, therefore repelling each other more intensely. However, neither water nor \mepy{} is found between neighboring \bphm{} ions. Therefore, the dielectric mismatch of the microphases may be an incomplete explanation of the larger ion separation.

In the ternary solution the distribution of \mepy{} is peaked more closely to \bphm{} than that of \NA{} counterions, similarly to the \nabph{}/\mepy{} case. In the ternary solution, however, \nabph{} pairs are almost always dissociated because of the strong preference of \NA{} for the aqueous environment.

Clearly the combination of both solvents, water and \mepy{}, is needed for \nabph{} to form supramolecular structures, with the crucial steps being (1) the separation of \NA{} from its counterion due to solvation by water, and (2) the hydrotropic action of \mepy{} enabling the formation of a combined hydrophobic aggregate with \bphm{}, whose shape is modulated by both the salt concentration and the \mepy{} volume fraction.

\subsection{Morphology of \nabph{}/\mepy{} aggregates}
\begin{figure}
	\begin{center}
  \includegraphics[width=0.9\columnwidth]{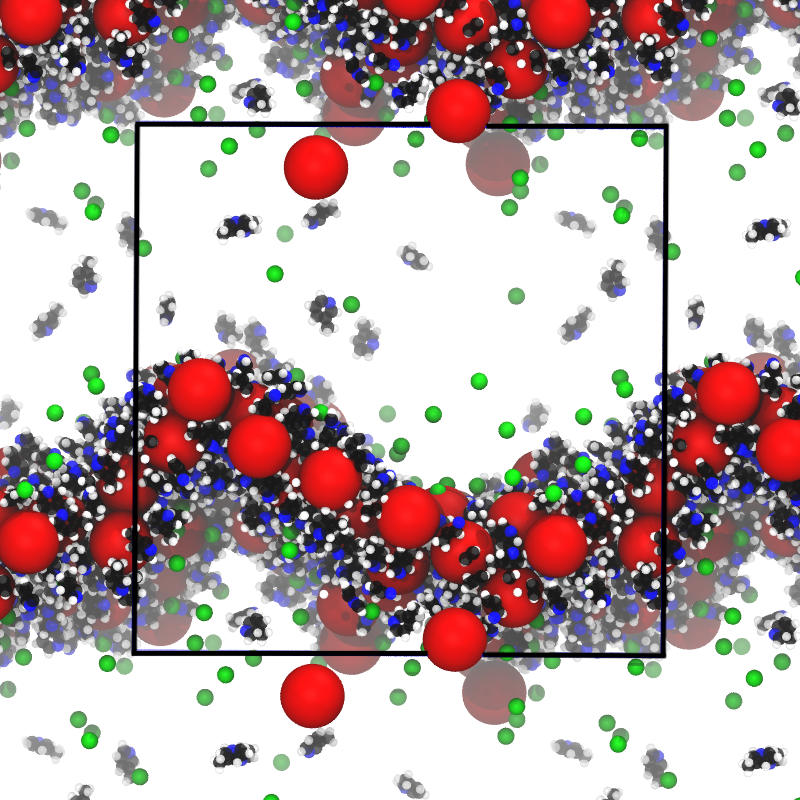}
  \includegraphics[width=0.9\columnwidth]{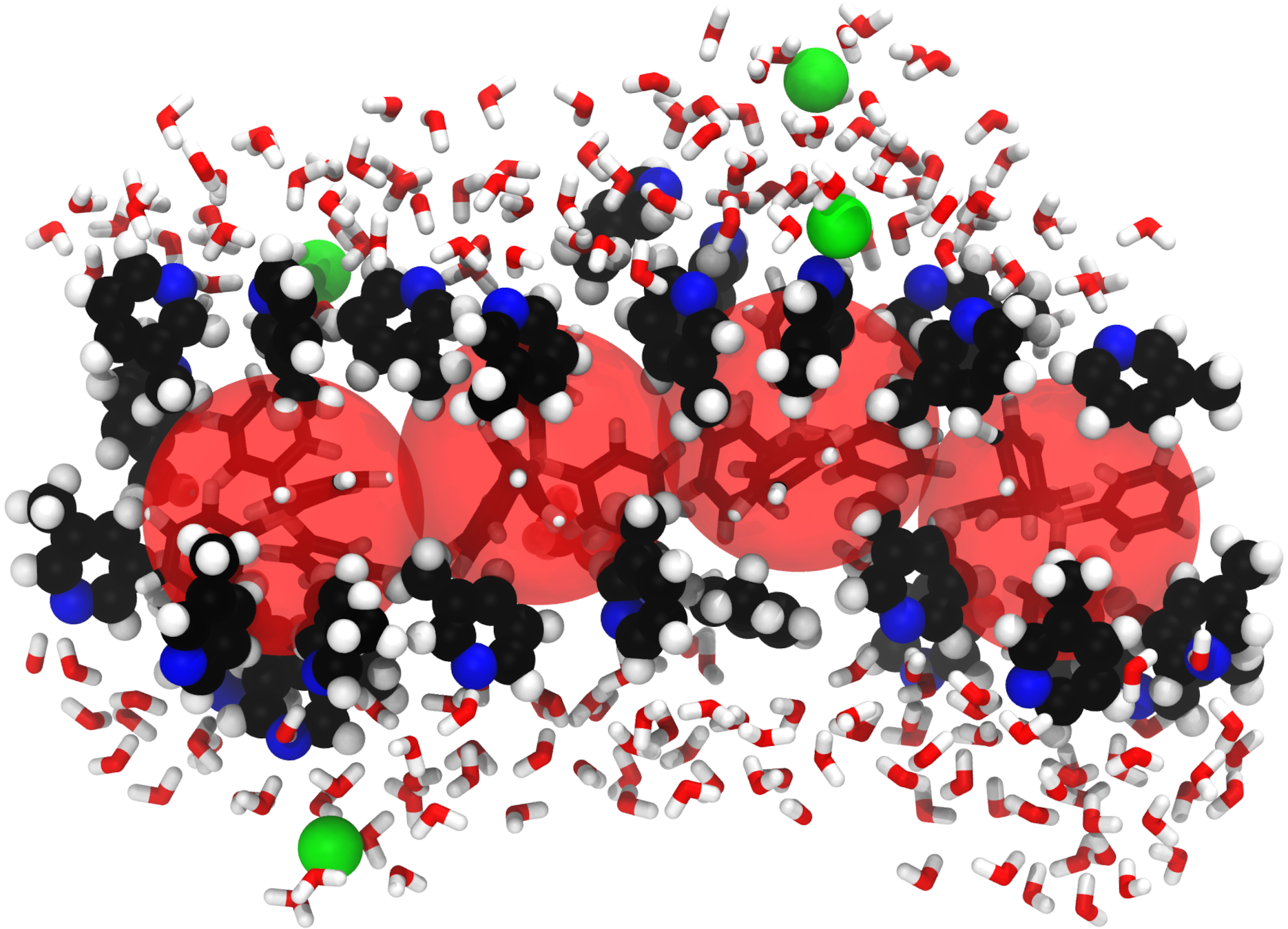}
  \end{center}
  \caption{Leaflet cross section at 15~vol\% \mepy{} and 300~mM salt (top: bead model, bottom: atomistic model, detail) showing \bphm{} (red spheres, sticks plus red halo), \NA{} (green spheres), \mepy{} (ball-and-stick representation), and water (sticks). Water molecules in the top panel have been removed for clarity. The black lines mark the simulation box.}
\label{fig:profile}
\end{figure}

Across a broad range of salt and \mepy{} concentrations, we found that mesostructures in ternary mixtures consist of the same building block exemplified in Fig.~\ref{fig:profile}, namely \bphm{} leaflets enveloped in one layer of \mepy{}. The leaflets deform strongly under thermal fluctuations and sometimes form transient vesicles when these fluctuations bring opposing ends of a leaflet into contact, occasionally enveloping one or more \NA{} ions, which could reduce the electrostatic self-repulsion of the \bphm{} anions. 
An example of such a vesicle is shown in Fig.~S1 in the SI.
The structures observed in the 8~nm edge length systems were also present in larger simulations (edge length 20~nm) that we performed on the JUWELS supercomputing cluster\cite{JUWELS} and used to test for finite-size effects.
Results from these larger simulations are likewise shown in the SI in Fig.~S2.

As a general rule, smaller concentrations lead to the formation of smaller leaflets.
Already at salt concentrations of 210~mM, leaflets span the entire simulation box even in the 20~nm systems.

We occasionally observed stable holes, about 5 \bphm{} in size, in some of the lamellar structures both in the small and in the large simulation box. This suggests that the observed holes are not an artifact of the periodic boundaries, but a genuine feature. Tasios and coworkers observed similar structures in their lattice-based coarse-grained Monte Carlo simulations of antagonistic salt mixtures.\cite{tasios_microphase_2017} The authors found the structure factor of these perforated leaflets to be practically indistinguishable from that of unperforated leaflets, rendering both types of structures plausible explanations of the experimentally observed ordered lamellar structures\cite{sadakane_membrane_2013}.

The leaflet structure becomes less discernible at high salt concentrations, as the \bphm{} anions form complicated three-dimensionally branched structures reminiscent of a sponge phase.
Still, the aggregates can be described as strongly curved interconnected leaflets or chains with thicknesses not exceeding one or two \bphm{} anions.

These structures would explain the experimental observations made by Sadakane and coworkers at salt concentrations larger than 300~mM, where the authors found a single peak in the SANS structure factor suggesting a sponge phase with a lamellar width of about 1.6~nm, that is, about one and a half times the diameter of \bphm{}. Unlike in the patterns measured at lower concentrations, no secondary Bragg peaks were detected at concentrations larger than 300~mM, indicating the disappearance of periodic lamellar arrangements\cite{sadakane_membrane_2013}.

As already mentioned, lamellae at all concentrations consist of a leaflet of \bphm{} anions enveloped by a thin layer of \mepy{}, with the overwhelming majority of \NA{} cations solvated in the water phase. As a result, the lamellae are negatively charged with a surface charge density that can be estimated from the local arrangement of \bphm{} ions to be about 1.0~$e$/nm$^2$, with $e$ the electron charge.

It is worth noting that one key ingredient in the leaflet formation is the size of the \NA{} counterions. If one uses \NA{} counterions with a Lennard-Jones energy parameter 20 times smaller than the correct one, the counterions' reduced effective size allows them to fit, together with their first water hydration shell, between the phenyl rings of \bphm{} ion pairs. As soon as \NA{} ions infiltrate the \bphm{} leaflets, the electrostatic self-repulsion no longer limits the growth of the \bphm{} clusters in volume, and spherical aggregates are formed in place of leaflets. An example of such a spherical aggregate is shown in Fig.~S3. When, on the other hand, there is no room in the hydrophobic \bphm{}/\mepy{} clusters for \NA{} ions and their solvation shells, \NA{} counterions remain in the aqueous phase. The aggregates try to maximize their interfacial surface area to minimize the electrostatic energy cost associated with a spatial separation of the ionic charges, eventually leading to the formation of leaflets.

From these results, one can appreciate the complex and delicate interplay between the various species' solvation forces, inducing aggregation of the otherwise mixing \mepy{} in water as soon as it is combined with \nabph{}. In turn, \nabph{} in pure \mepy{} does not aggregate because \mepy{} is not polar enough to trigger the hydrophobic effect. In pure water, \nabph{} would be close to the size limit of forming hydrophobic aggregates if it were neutral, even in the absence of dispersion forces\cite{chandler_interfaces_2005}. However, in this case, the electrostatic interaction provides sufficient repulsion to prevent aggregation beyond the formation of pairs or triplets at all but the highest salt concentrations.

\subsection{Bead model of \bphm{}}
To investigate the role of the hydrophobic effect in driving the aggregation, we repeated our simulations using a simplified bead model of \nabph{}. In this model, we represent the complex \bphm{} ion with a charged, pointlike particle, interacting with other \bphm{} ions via Coulomb and purely repulsive Lennard-Jones potentials only. In this way, we aim to exclude the contribution from attractive dispersion forces between \bphm{} ions and any influence from their molecular structure.

\begin{figure*}
\includegraphics[width=0.3\textwidth]{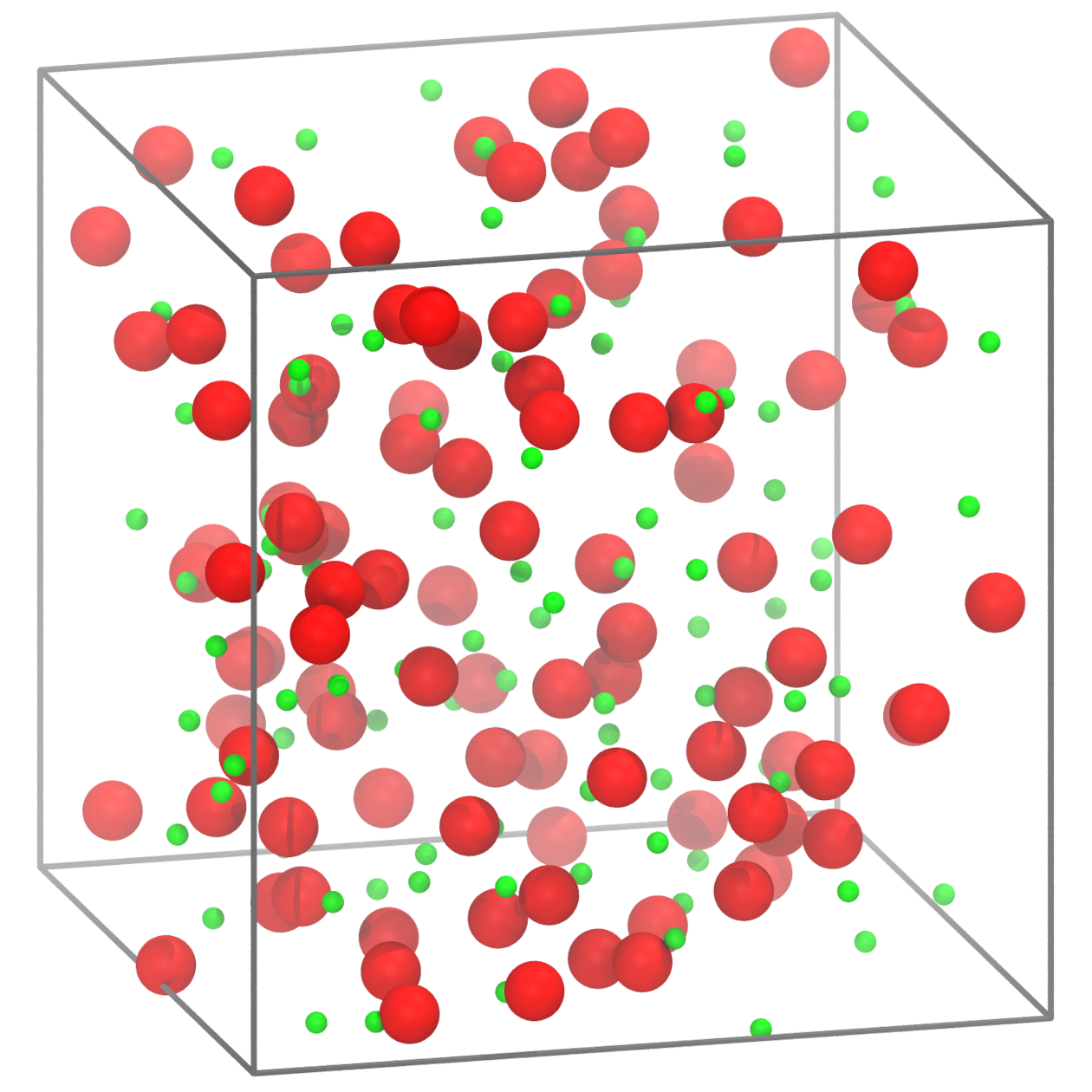}
\includegraphics[width=0.3\textwidth]{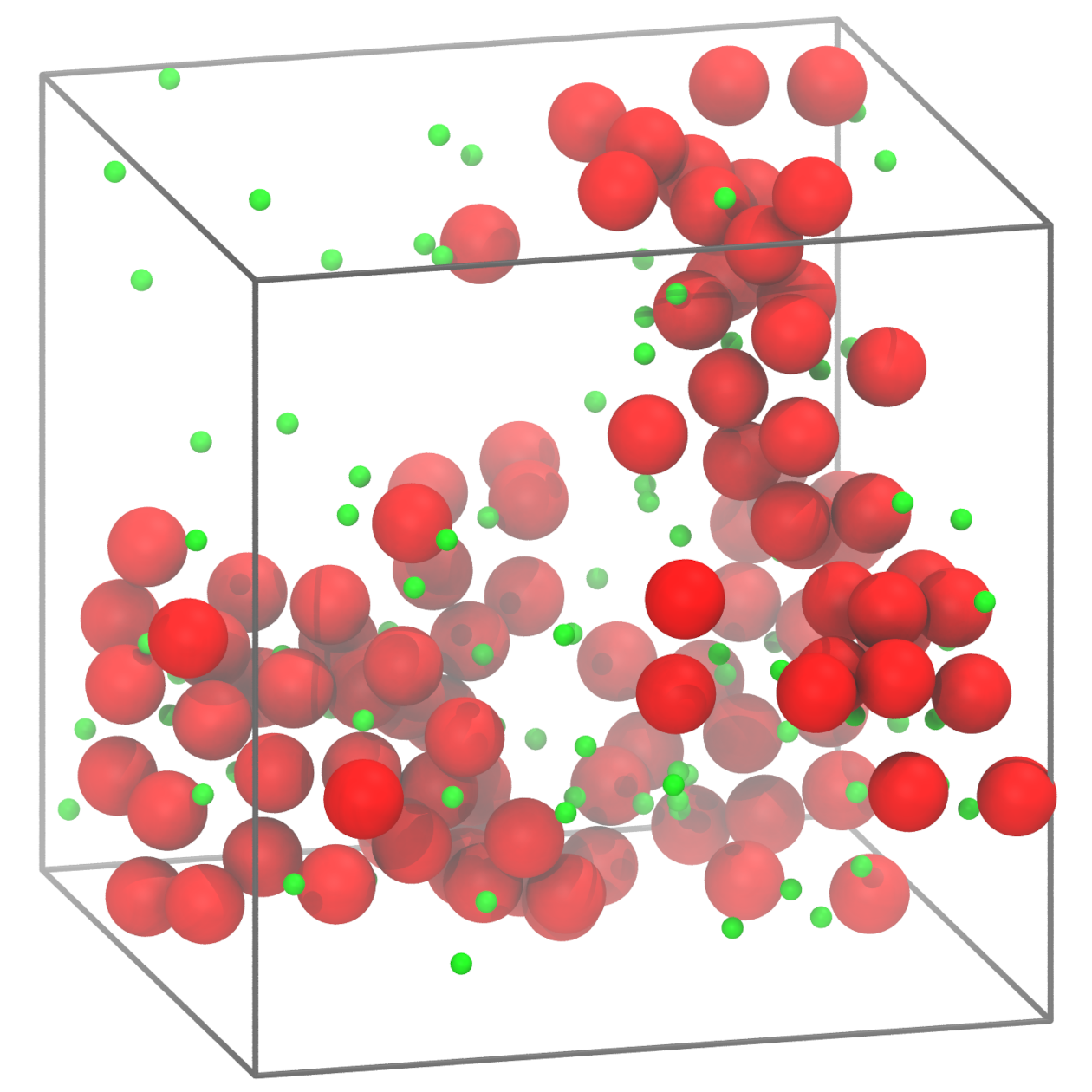}
\includegraphics[width=0.3\textwidth]{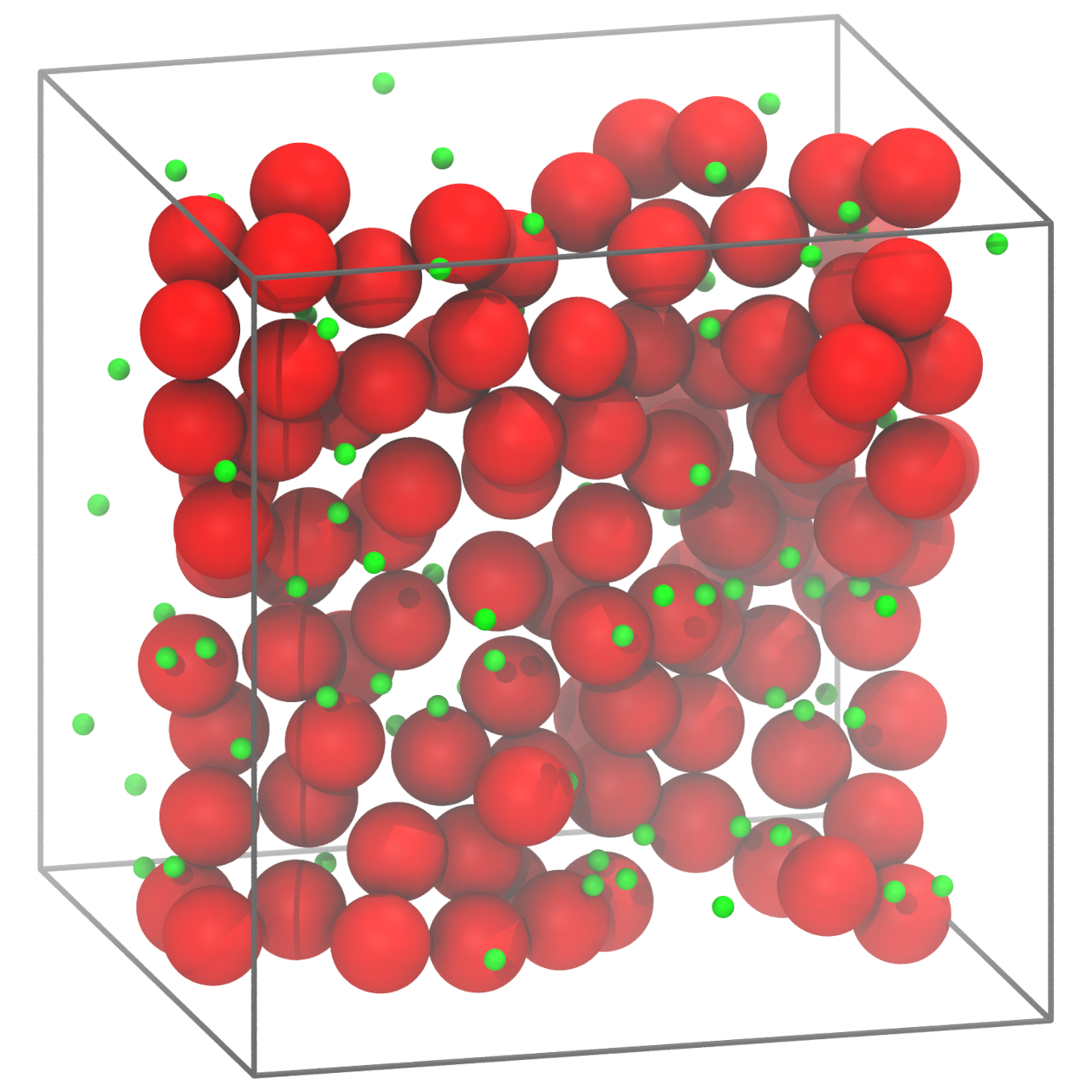}
\caption{Snapshots from bead model simulations at 15\% \mepy{} and 300~mM salt with large bead diameter $\sigma=0.6$~nm (left) and $\sigma=0.8$~nm (center), and $\sigma =1$~nm (right). Anions (red) and cations (green) both shown to scale.}
\label{fig:sigl}
\end{figure*}

In this sort of partial coarse-graining approach, each \bphm{} bead is assigned a charge of $0.85$ $e$, a Lennard-Jones potential with energy parameter $\epsilon$=1.47~kJ/mol (the same value from the Madrid model we used in the atomistic model for \NA{}), and a Lennard-Jones diameter $\sigma$=1~nm. The other ions and molecules were modeled in the same way as in the previous simulations. In the case of this bead model, however, we truncated the Lennard-Jones interaction at the minimum of the potential (at a distance of $2^{1/6}\simeq1.122$~nm),
thus implementing the purely repulsive Weeks-Chandler-Andersen potential\cite{weeks1971role} for the coarse-grained \bphm{} ions. Since the Lennard-Jones diameters of all other atoms are much smaller than 1~nm, all other interactions can be considered by all practical means the same as in the fully atomistic model.

As we demonstrate for three concentrations in Fig.~\ref{fig:bead_sample}, the structures we obtain with this simple bead model are almost identical to those obtained with the fully atomistic \nabph{} one. This shows that neither specific \bphm{}--\bphm{} attractive interaction, nor some kind of interlocking of the phenyl rings is needed to obtain the monomolecular \bphm{} layers. It is noteworthy that the size of the \bphm{} anions (1~nm) lies exactly at the crossover between entropy- and enthalpy-dominated regimes of the solvation free energy.\cite{chandler_interfaces_2005} Stable or metastable aggregates can form spontaneously (in water) only when a cluster exceeds the size threshold of 1~nm. Therefore, \bphm{} ions can directly act as hydrophobic building blocks for the monomolecular lamellae and do not first need to form primary aggregates that would in turn assemble into a thicker aggregate. This view is confirmed, as we show in Fig.~\ref{fig:sigl}, by the fact that even a slight decrease of the \bphm{} bead size to $\sigma$=0.8~nm strongly disrupts the leaflet shape. Upon a further decrease to $\sigma$=0.6~nm, \bphm{} aggregation practically ceases completely.

This strong dependence of structure formation on the size of the larger ion is consistent with experimental results from small-angle X-ray scattering by Witala and coworkers using mixtures of water, dimethylpyridine and 10~mM of tetra-n-alkylammonium bromide salt, whose cation was varied in size via the length of hydrocarbon chains attached to it.
By fitting a structure factor based on theory results from Onuki and Kitamura\cite{onuki_solvation_2004} to their scattering spectra, they observed that the coupling of the ions to opposite solvent phases becomes sufficiently strong to enable the formation of periodic structures around a size contrast slightly smaller than that between sodium and tetraphenylborate.\cite{witala_mesoscale_2016}

\subsection{Accessible surface area}
To quantify how well the structures conform to a leaflet shape and how accurately the atomistic and bead model results coincide, we performed measurements of the accessible molecular surface area (ASA)\cite{lee_interpretation_1971} as implemented in the freely available \texttt{pytim} analysis package\cite{sega2018pytim}.
By comparing the measured ASA values with those calculated from two simple models (planar leaflets and spherical micelles), we obtain a useful gauge to quantify whether our simplified picture of the molecular arrangement can or cannot be considered a fitting description. In these models we assume the \bphm{} anions to be beads with a radius $R$ = 0.5~nm arranged in hexagonal close packing on a plane or in a sphere.

In a first step we numerically calculated the average ASA per particle as a function of the number $N$ of aggregated particles in either planar or spherical close packing.
For this purpose we placed particles in either a 2D or 3D hexagonal lattice and then selected the closest $N$ particles around some arbitrary point in the lattice.
The obtained ASA as a function of $N$ is described well by simple power law fits, as shown in Fig.~S4.

\begin{figure}
  \centering
  \includegraphics{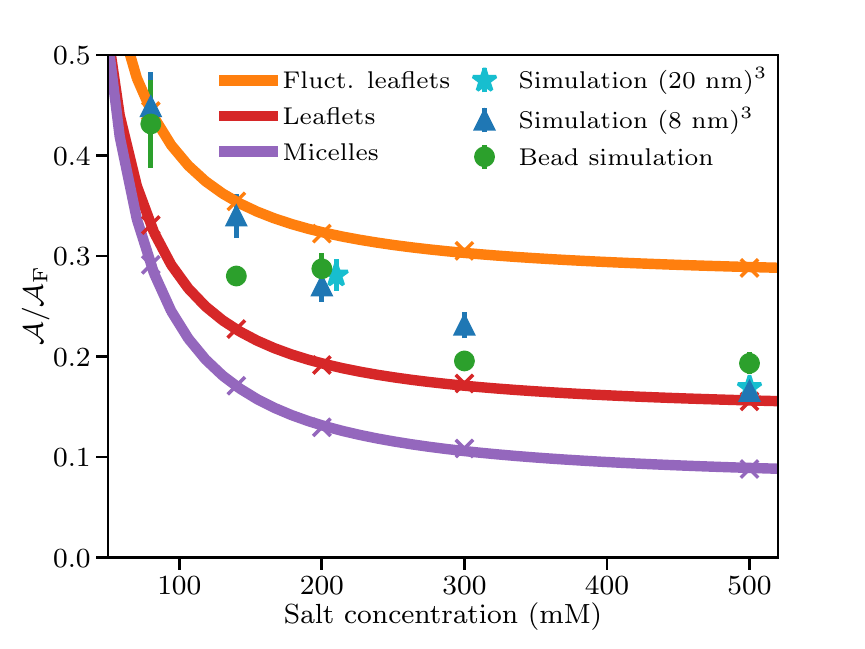}
  \caption{ASA $\mathcal{A}$ per \bphm{} accessible to a test sphere of 0.5~nm radius, as a function of \nabph{} concentration. Normalized by full surface $\mathcal{A}_{\mathrm{F}}$ of a 0.5~nm \bphm{} bead. All simulation results from trajectories at 15~vol\% of \mepy. Error bars show standard deviation over 25~ns.}
 \label{fig:asaa}
\end{figure}
Fig.~\ref{fig:asaa} reports a comparison of the measured ASA for both the bead model and the atomistic simulations with the idealized models of planar leaflets and spherical micelles with the same cluster size distribution as in the simulations.
To make the measured ASA of the bead and fully atomistic models comparable, we measured in the atomistic simulations the ASA of spheres of 0.5~nm radius centered on each \bphm{} anion.

The bead and fully atomistic models yield very similar ASA values for all concentrations.

In the central panels of Fig.~\ref{fig:bead_sample} we can see that the lamella obtained in the bead model connects with itself completely through the periodic boundaries, leading to the absence of border particles and hence a smaller ASA, whereas the atomistic model lamella does not contain enough material to fill the larger diagonal cross-section of the simulation box.
While the size and shape of the simulation box can clearly lead to some degree of variation in the measured ASA, the good agreement between simulations in 8~nm systems and those in 20~nm systems suggests that larger system sizes do not systematically change the obtained results.

The theoretical values in the leaflet and spherical cluster models decrease with concentration because of the increasing size of the clusters and a corresponding decrease in the ratio of border/inner molecules.
The simple leaflet model predicts ASA values that lie below those obtained in simulation for all concentrations.
This deviation can be qualitatively explained by the presence of thermal fluctuations displacing individual \bphm{} beads perpendicular to the local leaflet plane and thus coarsening the leaflet surface.
The fluctuating leaflet model in Fig.~\ref{fig:asaa} shows the substantial increase in ASA induced by adding a normally distributed displacement perpendicular to the plane with a standard deviation equal to one particle radius to each particle in the leaflet model.
While the amplitude of this displacement is chosen somewhat arbitrarily, the resulting model leaflet shapes look quite similar to the fluctuating leaflets obtained in simulations, as shown in Fig.~S4.
Besides the influence of thermal fluctuations, the ASA is also increased at small salt concentration due to a non-circular, frayed shape of the leaflets (compare Fig.~\ref{fig:bead_sample}).

The good agreement between the leaflet model and the measured ASA at high salt concentration in Fig.~\ref{fig:asaa} does not mean that the leaflets here are unperturbed by fluctuations.
Beyond about 200~mM of \nabph{}, the leaflets in both the 8~nm and 20~nm systems start connecting with their periodic images, becoming effectively infinite objects with no exposed boundary particles and, therefore, a lower ASA.
This decrease in the number of border particles is not reproduced in the leaflet model, which does not take periodic boundary conditions into account.
Furthermore, large salt concentrations lead to an increased frequency of local imperfections in coplanar particle arrangement due to crowding.

The average number of neighboring \bphm{} anions surrounding each \bphm{} within a 1.25~nm cutoff is $4.0\pm 1.4$, $4.8 \pm 1.3$, $5.4\pm 1.2$, $5.7\pm 1.4$, and $6.3 \pm 1.7$  at 80, 140, 200, 300, and 500~mM \bphm{}, respectively. One has to consider that the fraction of \bphm{} at the border of the aggregates (thus with fewer neighbors than 6) is higher at lower concentrations, thus decreasing the average number of neighbors.
The fraction of \bphm{} in the system with a number of neighbors between 3 and 6, which can be considered consistent with a planar hexagonal arrangement taking into account border particles, is $79\%$ (80~mM), $89\%$ (140~mM), $80\%$ (200~mM), $70\%$ (300~mM), and $56\%$ (500~mM).
At high salt concentrations, a growing fraction of particles has 7 or 8 neighbors due to surface-adsorbed additional \bphm{} particles or intersections between distinct leaflets in the sponge phase.

In summary, the measured ASA values and cluster statistics confirm the picture suggested by visual inspection, of a locally planar particle arrangement perturbed by thermal fluctuations and crowding defects in a relatively broad range of concentrations.
\begin{figure}
  \centering
  \includegraphics{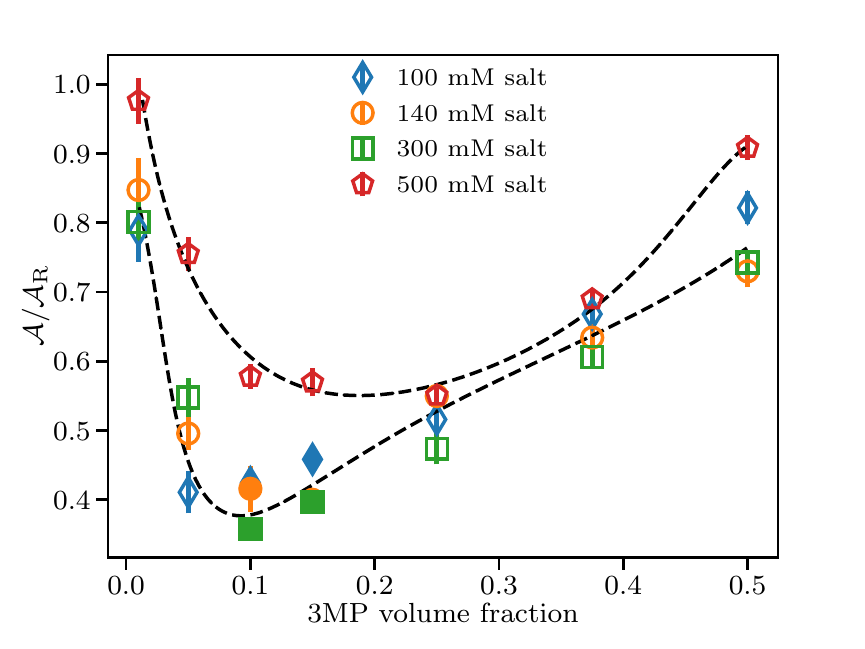}
  \caption{ASA $\mathcal{A}$ per \bphm{} accessible to a test sphere of 0.5~nm radius, as a function of \mepy{} volume fraction. Results normalized by expected ASA $\mathcal{A}_{\mathrm{R}}$ under random particle distribution. The dashed lines are spline fits and serve only as guides to the eye. Filled symbols mark lamellar phase structures. Error bars show standard deviation over 25~ns.}
  \label{fig:asab}
\end{figure}

To interpret the ASA results further, we found it helpful to introduce the concept of the ASA $\mathcal{A}_{\mathrm{R}}$ of a random distribution of \bphm{} beads. In practice, we determined the function $\mathcal{A}_{\mathrm{R}}$ as the best fit of a power law (similarly to the case of the leaflet and micelle models) to the ASA values of randomly placed non-overlapping spheres in 3D.

The ASA at salt concentrations up to 300~mM seems to decrease following about the same functional form as $\mathcal{A}_{\mathrm{R}}$. In fact, in Fig.~\ref{fig:asab}, $\mathcal{A} / \mathcal{A}_{\mathrm{R}}$ is only weakly dependent on the salt concentration, except for the 500 mM case.
Even though $\mathcal{A}$ is smaller at 500~mM than at lower salt concentrations, it is comparatively closer to the ASA $\mathcal{A}_{\mathrm{R}}$ of a random distribution. Correspondingly, $\mathcal{A} / \mathcal{A}_{\mathrm{R}}$ is the largest at 500~mM.
Dense clustering of \bphm{} eventually becomes unavoidable for purely entropic reasons as the salt volume fraction increases, and $\mathcal{A} / \mathcal{A}_{\mathrm{R}} \approx 1$ is to be expected regardless of interparticle interactions when \bphm{} concentrations approach close packing.
At 500~mM, the volume fraction of \bphm{}, represented by 0.5~nm beads, reaches about 16\%.
Crowding at high salt concentrations apparently induces a particle arrangement slightly more similar to random placement instead of the entropically costly generation of separate lamellae.

Small values of $\mathcal{A} / \mathcal{A}_{\mathrm{R}} \ll 1$ indicate that aggregation is enhanced by effective interparticle interactions rather than occuring merely due to volume constraints.
With this interpretation in mind, Fig.~\ref{fig:asab} suggests that \mepy{} concentrations in the range of 10-15 vol-\%, where the lamellar phase is observed, yield ideal conditions for the hydrophobic aggregation of \bphm{}.
Both at larger and smaller \mepy{} volume fractions \bphm{} tends to cluster less densely, leaving a larger ASA.
\subsection{Phase diagram}
\begin{figure*}
  \begin{tabular}{rl}
  \includegraphics{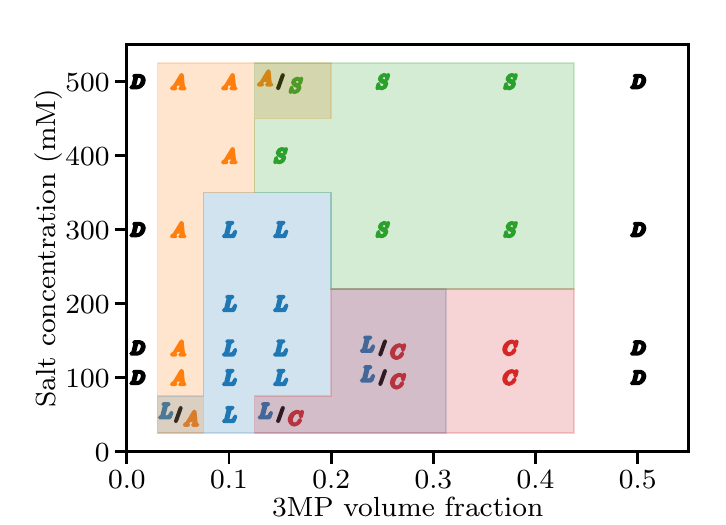} &
  \includegraphics[width=0.3\textwidth]{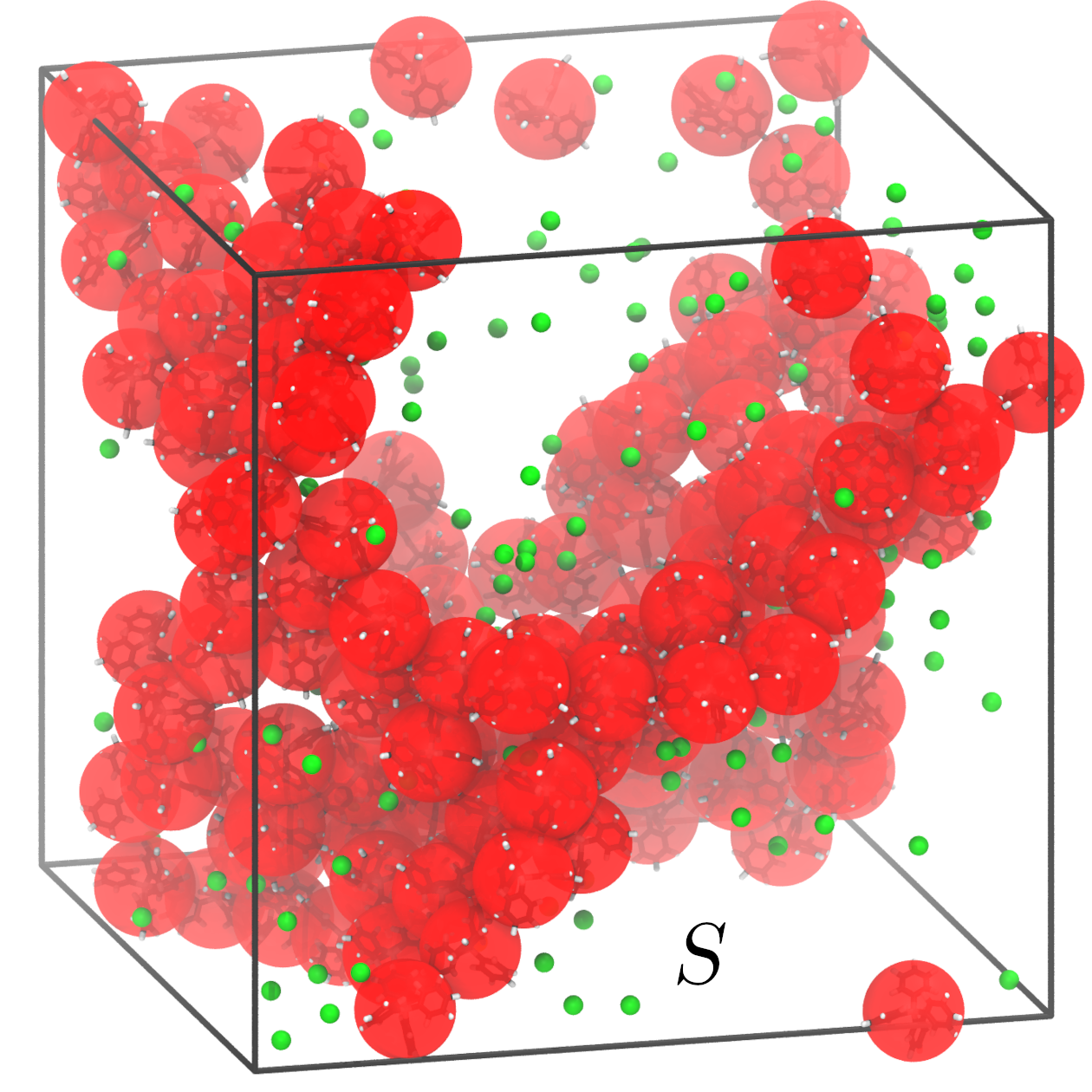} \\
  \end{tabular}
  \includegraphics[width=0.3\textwidth]{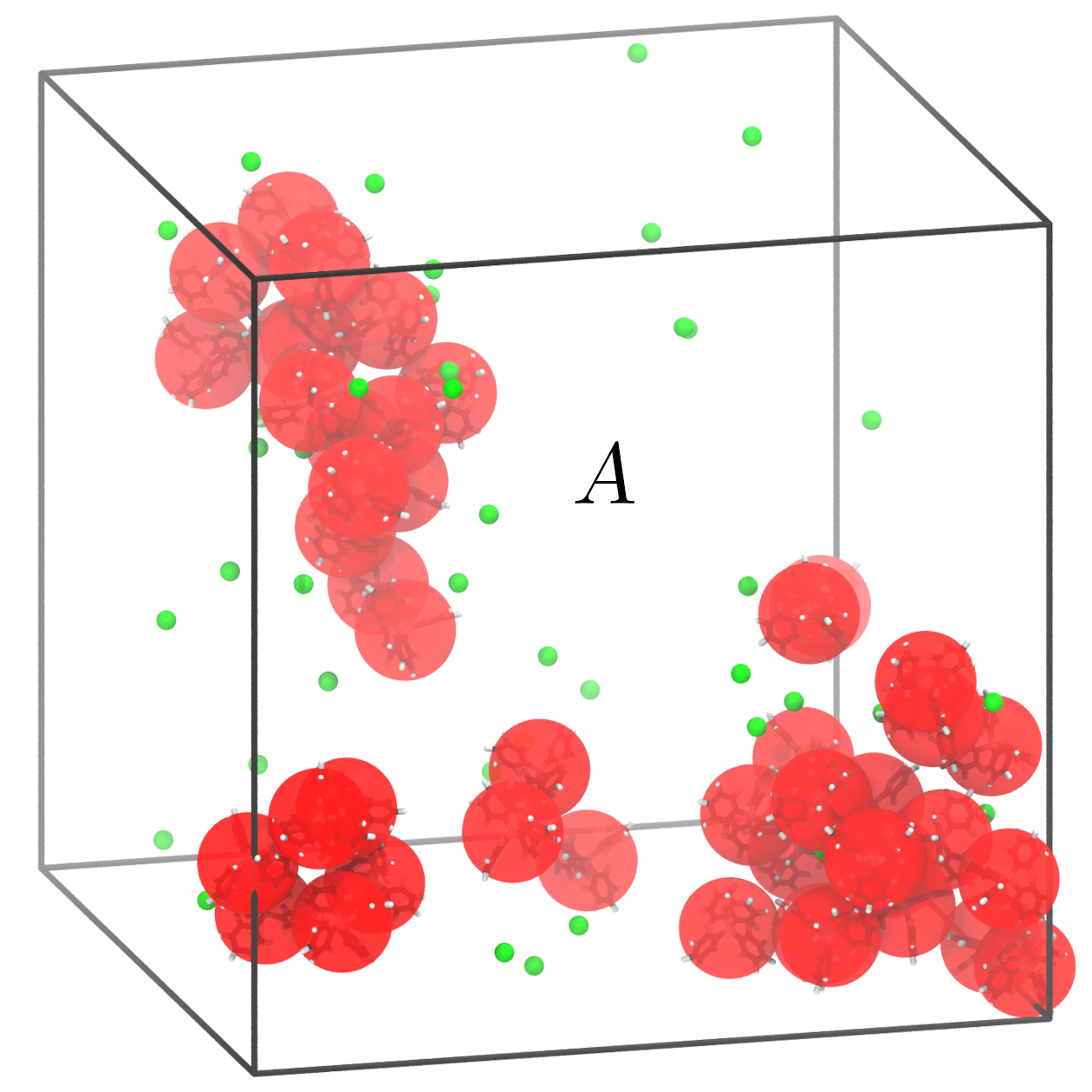}
  \includegraphics[width=0.3\textwidth]{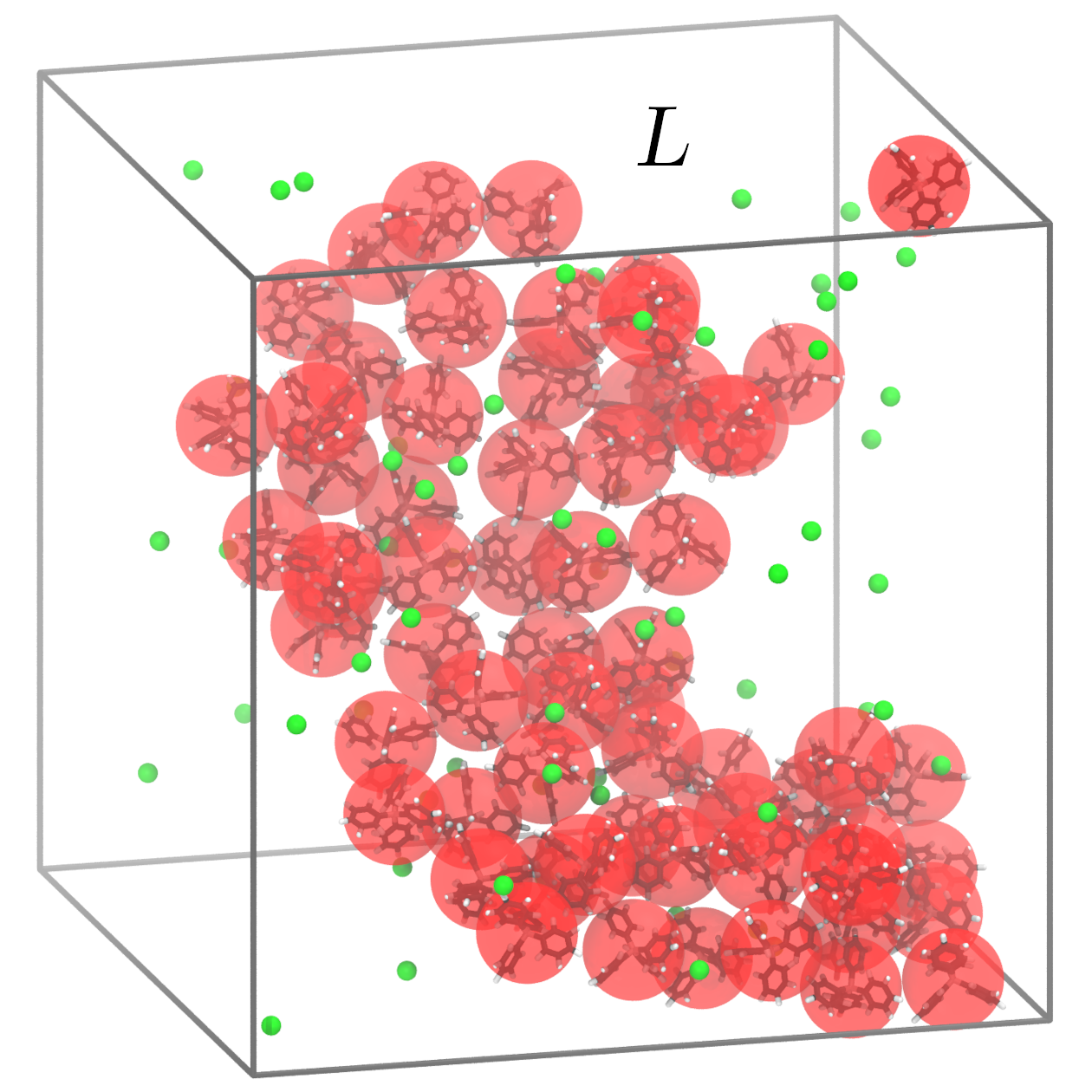}
  \includegraphics[width=0.3\textwidth]{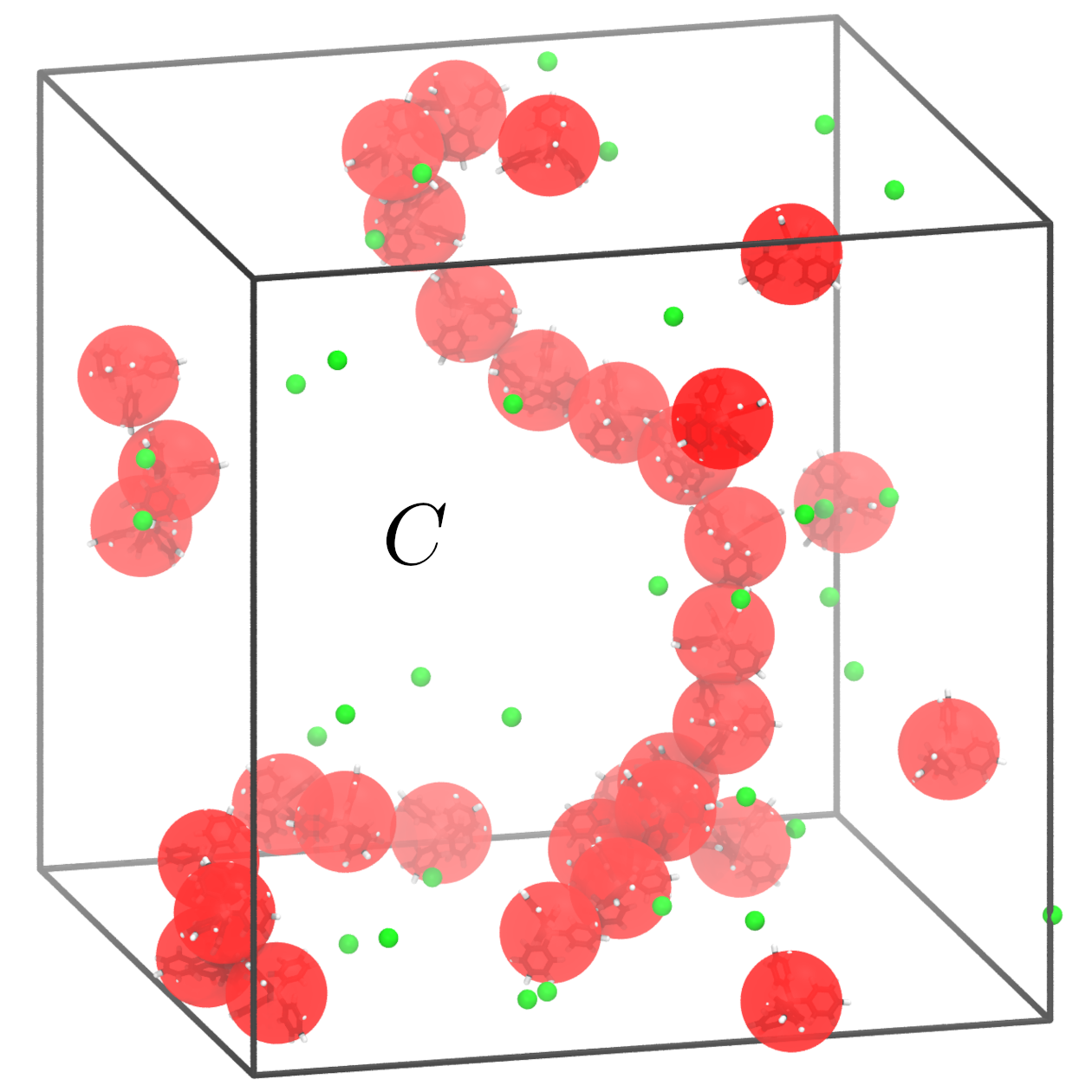}
  \caption{Phase diagram of \nabph{}/\mepy{}/water mixture interpreted from atomistic 8~nm simulations. $D$ stands for disordered arrangements, where \bphm{} is mostly randomly distributed, $S$ stands for the sponge phase (see {upper right} pane for 25\% \mepy{} and 500~mM salt), consisting of 3D interconnected leaflets and chains, $C$ stands for chains (see {lower right} pane for 37.5\% \mepy{} and 100~mM salt), $L$ stands for the lamellar phase, i.e. planar leaflets (see {lower center} pane for 10\% \mepy{} and 200~mM salt), and $A$ stands for amorphous aggregates (see {lower left} pane for 5\% \mepy{} and 140~mM salt).}
  \label{fig:phasediag}
\end{figure*}
Fig.~\ref{fig:phasediag} presents a phase diagram that summarizes our findings.
As indicated in Fig.~\ref{fig:asab}, both small and large \mepy{} concentrations are detrimental to the aggregation of \bphm{} , so that only disordered arrangements of dissolved \bphm{} anions are observed at both extremes of \mepy{} concentration.
Approaching the limit of \nabph{} solvation in pure water, \bphm{} forms only small, transient clusters because the electrostatic self-repulsion of \bphm{} impedes the growth of large hydrophobic aggregates without the assistance of the moderately hydrophobic \mepy{}.
When there is a surplus of \mepy{} in the system, on the other hand, the lack of water makes the formation of hydrophobic \bphm{} clusters unneccessary.
\bphm{} can loosely dissolve in \mepy{}-rich microphases without discernible ordering, avoiding the entropic and electrostatic energy costs associated with a compact arrangement of \bphm{} anions.

In Fig.~\ref{fig:phasediag}, we label any structure as belonging to the sponge phase that connects to itself through periodic boundaries in all three dimensions and consists of intersecting leaflets and/or chains with a thickness not regularly exceeding one \bphm{}.
The sponge phase consistently exhibits a monomolecular leaflet microstructure at higher \mepy{} concentrations, as in the example shown in Fig.~\ref{fig:phasediag}.
Note that any seemingly isolated \bphm{} particles in the example shown in Fig.~\ref{fig:phasediag} form part of a leaflet structure when the periodic boundaries are taken into account.

Just like the lamellar phase, the chain phase that  occurs at large \mepy{} but small salt concentrations turns into the sponge phase at higher salt concentrations.
Compared to the lamellar phase, the chain phase turns into the sponge phase at slightly lower salt concentrations.
This can be understood as a consequence of the decreased pressure for the \bphm{} aggregates to limit their surface area in order to avoid the water phase when there is little water in the system.

To the best of our knowledge the chain phase has not been yet identified in experiments. The majority of experimental studies on \nabph{} mesophases so far have focused on the \mepy{} concentration range in which the lamellar phase is observable, due to its characteristic Bragg peaks visible in SANS scattering.
Sadakane et al. performed some SANS measurements in a parameter range corresponding to the appearance of chain structures in our simulations, but, unfortunately, they could not extract any details about the microstructure of the observed phases apart from a characteristic periodicity of about 6 to 10~nm~\cite{sadakane_periodic_2007}, leaving the question of the existence of this phase open.

The lamellar phase has been discussed in detail in the previous sections.
Fig.~\ref{fig:phasediag} shows an additional example of a lamella exhibiting a large degree of bending due to thermal fluctuations.
At different stages of the simulation this bending almost caused the lamella to wrap around and form a spherical, vesicle-like shape comparable to Fig.~S1.
A few deviations from a perfectly monomolecular leaflet are visible near the bottom of the illustration, where individual \bphm{} particles have aggregated on top of the lamella.

The phase of amorphous aggregates is characterized by structures without a particular geometry, with a thickness often exceeding one \bphm{}.
When \mepy{} is scarce in the system, the formation of hydrophobic clusters with \bphm{} can still be observed, but \mepy{} is not sufficiently available to envelop a planar \bphm{} leaflet, forcing the \bphm{} aggregates to form thicker aggregates despite their electrostatic repulsion.

The transition regions between the phases are occasionally ambiguous.
Simulations containing either several aggregates pertaining to different phases or single aggregates deforming in time to resemble either one phase or the other were labeled as pertaining to both phases.
At salt concentrations below about 100~mM the average size of clusters decreases, making it progressively more difficult to distinguish the different phases.
\section{Conclusions}
We showed that monomolecular leaflets of \bphm{} appear in ternary \nabph{}/\mepy{}/water mixtures, forming lamellae stabilized by a substantial surface charge when the salt concentration is up to 300~mM and the \mepy{} concentration is about 10-15~vol\%. The leaflets consist of one layer of \bphm{} ions enveloped by one molecular layer of \mepy{} on either side and a cloud of \NA{} counterions solvated by water.

The leaflet thickness, including both \bphm{} and \mepy{}, can be estimated from the boron-nitrogen radial distribution functions to be approximately 1.6~nm.
These observations agree with the presence of lamellae with a thickness of about 1.5$\pm$ 0.5~nm at similar salt and \mepy{} concentrations suggested by SANS data\cite{sadakane_membrane_2013}.  Our simulations suggest that lamellae form if enough \mepy{} is available to envelop the \bphm{} leaflets, but not as much as to provide a large surplus of \mepy{} in the surrounding solvent, as this would encourage \bphm{} to form aggregates with a larger surface to volume ratio, such as chains of individual \bphm{}. Also, the \bphm{} concentration should be large enough to allow for the aggregation of enough \bphm{} to form a large leaflet, but not as much to reach the sponge phase.

We have also shown that practically identical structures can be obtained by replacing the fully atomistic model of \bphm{} ions with a simpler model of charged, repulsive spheres. The leaflet formation and stabilization can be explained simply by considering the ionic charge and the hydrophobic effect, without the need for attractive dispersion forces between the \bphm{} ions. 

In this sense, we expect that mixtures with similar properties could be realized by employing as three main ingredients (1) a salt with large size difference between anion and cation, with one of the two larger than the 1~nm critical threshold for hydrophobic aggregation (2) a moderately polar co-solvent exhibiting hydrotropic features and (3) a polar solvent like water.

\section{Supporting information}
Includes additional figures depicting larger simulations and structural details, as well as links to our simulation data and topologies used.
\section{Acknowledgments}
The authors gratefully acknowledge the Gauss Centre for Supercomputing e.V. (\texttt{www.gauss-centre.eu}) for funding this project by providing computing time on the GCS Supercomputer JUWELS at J{\"u}lich Supercomputing Centre (JSC).
\bibliography{main}
\clearpage
\begin{figure}
  \includegraphics{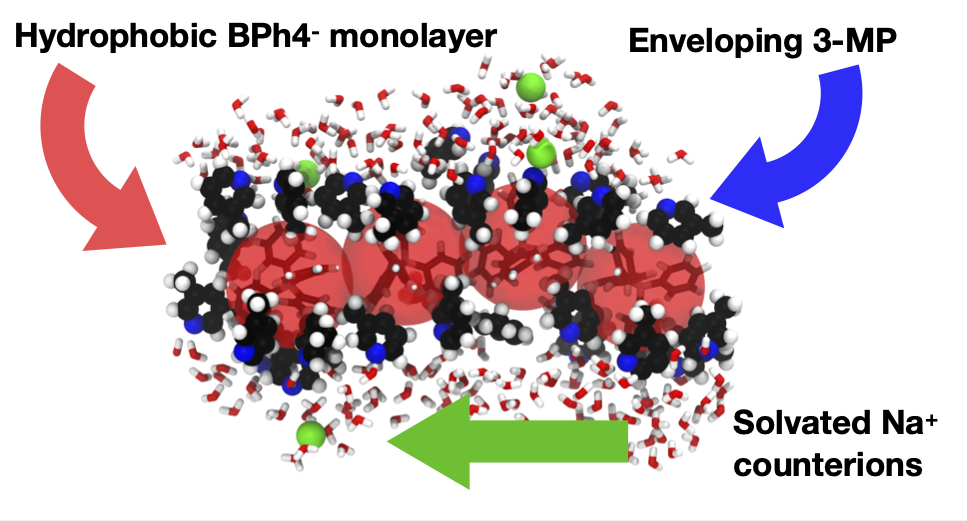}
  \caption{For Table of Contents Only}
\end{figure}

\end{document}